\documentclass[a4paper,11pt]{article}
\usepackage{pos}

\title{Asymptotic Safety and the Cosmic Coincidence Problem}

\author*{Vasilios Zarikas}

\affiliation{Department of Mathematics, University of Thessaly,\\3rd Km Lamia Athens, Lamia, Greece}

\emailAdd{vzarikas@uth.gr}

\abstract{Asymptotic Safety (AS) Paradigm is an interesting set of ideas and methods towards a meaningful quantization of Gravity. A brief review of phenomenological consequences in the context of AS regarding cosmology with emphasis on dark energy is given. Furthermore, recent studies that suggest a natural explanation of the recent cosmic acceleration and its coincidence using large-scale structure and AS framework will be analyzed. Finally, the present work extends the analysis of a recent new set of modified Einstein equations inspired from AS program. }

\FullConference{%
  Corfu Summer Institute 2022 "School and Workshops on Elementary Particle Physics and Gravity",\\
  28 August - 1 October, 2022\\
  Corfu, Greece}

\begin{document}
\maketitle

\section{Introduction}

There are several promising attempts towards a theory of quantum gravity; all of them still not satisfactory. The most popular frameworks that try to address the description of gravity at the fundamental level are: String-branes theory / M theory,  Loop quantum gravity and most recently the  Asymptotic Safety program (and similar RG approaches).
String theory, by far the most popular and workable framework, operates in more than three spatial dimensions and assumes a continuum background that needs to be compactified. On the other hand Loop quantum gravity (or spin foam models) is a background independent methodology defined in 3+1 dimensions, handling discrete spectrum of quantum spacetime observables. It has certainly some conceptually attractive features.
Asymptotic safety (AS) ,\cite{ASreviews}, \cite{Falls:2014tra}, works also in 4-dim and with continuous manifolds. It is a minimal proposal that keeps the same symmetries and fields of Quantum Field Theory (i.e. SM) and General Relativity (GR). It was able to indicate that general relativity and other extensions of the Einstein Hilbert action (with higher derivatives) can be non-perturbatively renormalizable models. It is not a complete framework since we don’t know the Lagrangian of our world. Finally, it is a background independent approach. The steps that are followed in the framework are. First we define the theory meaning the field contents (e.g. graviton other gauge fields) plus symmetries (e.g. coordinate trns). We always respect diffeomorphism invariance plus some gauge symmetries from the non gravity sector like SM gauge symmetries. Next, we determine the family of actions i.e. specific interactions of fields that respect symmetries. Consequently, we work with functional renormalization group equations in the theory space. The space containing all actions with “coordinates” coupling constants (e.g. $G$, $\Lambda$), The renormalization group flow connects physics at different scales $k$, that’s why at the end we have $G$, $\Lambda$ as running coupling constants.

In conventional quantum field theory quantum systems are treated with many degrees of freedom and many orders of magnitude in length, or in energy and momentum. However, the measurements are typically performed at low energies, in the infrared (IR) regime.  A different mathematical model/theory where those quantum fluctuations or modes are taken into account for an energy which is between the UV and IR energy scales is needed. The functional renormalization group (RG) method is such a mathematical tool to take into account the quantum fluctuations step by step, consistently.  
The ultraviolet Lagrangian contains interaction terms multiplied by the couplings. The renormalisation group flow method consistently eliminates the degrees of freedom of the theory as we go towards lower energy and describes the values of the running couplings, \cite{FRG}. Finally, we can obtain the value of IR couplings at practically zero energies. In the IR the effective potential usually has an involved structure. 

Weinberg first proposed in 1976 \cite{c11}.  that perturbative renormalizability is not the only way for a theory to be complete at high energies. It is enough to have finite values, at UV energies, to a finite number of parameters that appear in the action.  Asymptotic safety proposes just this thing, a theory to have these two constraints: 1) a finite number of finite parameters that determine the theory at high energies 2) with finite fixed values.
While the general idea has been around for many years, it has only been in the late 1990s, following works by Wetterich and Reuter, that asymptotically safe gravity has been formulated, \cite{c12}, \cite{FRG}.

In the present work we review the cosmological consequences of the gravity sector of an action that exhibits Asymptotic Safe behaviour. In this context, we name these modified gravity theories as Asymptotic Safe Gravity. There are several works that try to solve various different problems of High energy physics, Cosmology and Astrophysics using the properties of Asymptotic Safe Gravity. There are for example several attempts that try to explain inflation, dark energy, dark matter, to recover the MOND dynamics and Hubble tensions. Furthermore, results from a series of papers concerning AS swiss cheese model with an interesting solution of the dark energy problem will be given.  \cite{Zarikas:2017gfv}, 
\cite{Anagnostopoulos:2018jdq},
\cite{Anagnostopoulos:2019mrc},
\cite{Anagnostopoulos:2022pxa}, \cite{Mitra:2021ahd}.
Finally, focusing again on the dark energy model the low energy cosmic evolution of recently developed minimally modified Einstein equations compatible with AS gravity, \cite{Bonanno:2020qfu}, will be given for the first time.

\section{Cosmological implications of Asymptotic Safe Gravity}

Several interesting works have appeared in literature that study the cosmological and astrophysical consequences of AS gravity \cite{cosmoreview}. They can be categorised in three cases.
First, models that perform RG improvement implemented at the level of the equations of motion, which should be handled with care as far as the energy conservation is concerned (in the absence and in the presence of matter) and
second, studies that start working with an AS compatible effective action. Finally the most crude approach is the improvement on existing classical solutions of general relativity, which may be accepted for small time intervals of the cosmic evolution and not for the whole cosmic history range. 
These approaches are  approximations since AS framework although a rigor methodology capable of proving UV completeness for a class of actions, resembles the existence theorems of differential equations and thus cannot not provide or explain yet the fundamental Lagrangian or the emergence of the cosmological spacetime.
However, it is expected that these RG improvements provide a fair description of AS phenomenological consequences assuming of course that the AS framework is the correct description of Nature.

Interesting works that connect the horizon and flatness problems and the inflationary period within the AS framework are \cite{Bonanno:2001xi, Reuter:2005kb, Bonanno:2007wg, Weinberg:2009wa, Kofinas:2016lcz, Contillo:2010ju, Bonanno:2010bt, Hindmarsh:2011hx,Cai:2011kd,Contillo:2011ag, Cai:2012qi, Bonanno:2012jy, Hindmarsh:2012rc, Fang:2012ca, Copeland:2013vva,Xianyu:2014eba,Saltas:2015vsc,Nielsen:2015una,Bonanno:2015fga, Gubitosi:2018gsl, Lehners:2019ibe}. The idea is that the running of a positive cosmological constant $\Lambda$ which is treated as a coupling, can trigger the initial acceleration, while the exit from it, is guaranteed due to the RG flow that reduces the value of $\Lambda$. There is also the possibility to get inflation from higher derivative gravity actions that are treated in the AS framework and they have a non Gaussian fixed point, which means that they are non-perturbatively renormalised,  because they phenomenologically behave like $R2$ gravity. 
The cosmic entropy issue usually connected with the particle creation after inflation during the thermalisation has also been considered.
Studies that provide solutions to the cosmic entropy problem are \cite{Bonanno:2007wg, Bonanno:2010mk, Kofinas:2016lcz}

There are also several proposals that try to explain the so called dark energy problem with the RG evolution of $\Lambda$, see \cite{Bonanno:2001hi, Bentivegna:2003rr, Ahn:2011qt, Zarikas:2017gfv, Anagnostopoulos:2019mrc}. We will elaborate more on this in the next section.

Concerning the important issue of spacetime singularity avoidance/resolution, AS gravity can offer also solutions, \cite{Casadio:2010fw, Zarikas:2018wfv, Kofinas:2016lcz, Kofinas:2015sna, Bonanno:2017zen}. Quantum gravity approaches that assume a discrete spacetime or an initial network of events can resolve naturally the problem of spacetime singularities. AS is a theory that encapsulates a continuous gravity field. The reason that AS can provide singularity avoidance is the anti-screening of gravity strength  as well as the possible existence of a positive $\Lambda$ in the UV regime.

Asymptotic Safe Gravity modifies also the properties of black holes such as non singular center, inner horizon, possible final remnants and radiation, as it is suggested in \cite{Eichhorn:2022bgu, Falls:2010he, Cai:2010zh, Saueressig:2015xua}

\section{Swiss cheese and recent cosmic acceleration}

It is not known if there is at the far infrared limit, ($k \rightarrow 0$), a fixed point. The infrared behavior of the RG flow trajectories describing the running behavior of $G$ and $\Lambda$ cannot be trusted since the RG methodology stops to be valid as $\lambda_{k}$ reaches the value, $1/2$. However, the divergence of the beta functions happens near $k=H_{0}$. Thus, AS cannot predict the exact current value of $\Lambda$. It may either be very negligible solving the old cosmological problem or it can be close to the $\Lambda CDM$ model value. In the latter case AS can explain the recent cosmological constant problem but with fine tuning and without answering about the coincidence problem. However, AS may explain both old and new cosmological constant, as well as the coincidence problem without fine tuning with the AS swiss model first proposed in \cite{Zarikas:2017gfv} and further studied and generalised in \cite{Anagnostopoulos:2022pxa}, \cite{Mitra:2021ahd}, \cite{Anagnostopoulos:2019mrc}, \cite{Anagnostopoulos:2018jdq}.

The recent cosmic acceleration that is attributed to a Dark energy (DE) component of the Universe,\cite{Li:2011sd},  perhaps is the result of a time varying cosmological constant $\Lambda(z)$. We will present now, the proposed solution of \cite{Zarikas:2017gfv}, \cite{Anagnostopoulos:2022pxa}, based in AS theory.
It removes the coincidence and fine tuning problem connecting the recent large scale structure and its characteristics with the recent cosmic acceleration. There are also some other works, \cite{inho},
\cite{inhoswiss}, \cite{structure}, that resolve partially the recent cosmic acceleration and coincidence problem using the recent large scale structure formation; these works prove that inhomogeneities can generate up to a point a DE effect. In \cite{Zarikas:2017gfv},\cite{Anagnostopoulos:2022pxa},  the local value of a positive $\Lambda$ associated with galaxies or clusters of galaxies is the reason of the solution. 

AS swiss model,\cite{Zarikas:2017gfv}, generalised in \cite{Anagnostopoulos:2022pxa}, works with $G_{k}$ and $\Lambda_{k}$ not in the UV regime (near the NGFP fixed point) or in the far infrared scale, but in the intermediate astrophysical scale. Other studies that assume such IR corrections at astrophysical scales are, \cite{IR}.
	
The AS swiss cheese model also elaborates AS inspired quantum improved Schwarzschild-de Sitter metric Eq.(\ref{ASBH}), \cite{Koch:2014cqa}. These metrics have been chosen to describe the homogeneously distributed galaxies and/or clusters of galaxies.
	
AS theory proposes that $G_{k}$ and $\Lambda_{k}$ are functions of the energy scale related cutoff $k$. 
Note that the cutoff $k$ is not as a momentum flowing into a loop but can be related to an inverse of a typical 
distance over which we average the field. So, in a cosmological context of use, $k$ could be related to a cosmic time, \cite{c4}, or cosmic/astrophysical distance. 
In the AS swiss cheese the scale $k$ is connected with the typical size of a cluster of galaxies. 

The original swiss-cheese model or Einstein-Strauss model  \cite{Einstein:1946ev} describes a global homogeneous and isotropic Universe that consists locally of many homogeneously distributed Schwarzschild black holes. The spacetime matching of the global exterior metric with a local spherical black hole is achieved respecting the Israel-Darmois conditions \cite{Israel:1966rt}. A smooth matching of the two solutions happens if the first fundamental form (the intrinsic metric) is equal and the second fundamental form is opposite for the two metrics across the whole matching surface if such a surface exists.

The AS swiss cheese model matches an AS corrected
Schwarzschild-de Sitter metric with a homogeneous isotropic metric. Both metrics contain $k$ dependent cosmological and Newton constants.
Thus, we have
\begin{equation}
ds^2=-\Big(1-\frac{2G_k M}{R}-\frac{1}{3}\Lambda_k R^2\Big)dT^2
+\frac{dR^2}{1-\frac{2G_k M}{R}-\frac{1}{3}\Lambda_k R^2}
+R^2\,d\Omega^2\,,
\label{ASBH}
\end{equation}
where $G_k$ and $\Lambda_k$ are functions of a
characteristic scale $k$ of the system under investigation.
The homogeneous and isotropic metric is of the form
\begin{equation}
	ds^2=-dt^2+a^2(t)\left[\frac{dr^2}{1-\kappa r^2}+r^2d\Omega^2
	\right]\,,
	\label{eq:FRW}
\end{equation}
where $d\Omega^2=\left(d\theta^2+\sin^2\!\theta\,d\varphi^2\right)$ the metric of the two sphere and $a(t)$ is the scale factor and $\kappa=0, \pm 1$ the spatial curvature

The matching requirements provide the following equations for $R_{S}$ which is the value of black hole radius in the matching surface ($r=r_\Sigma$), which is related with the cosmic scale factor,
\begin{eqnarray}
R_{S}&=&ar_{\Sigma}\label{M1}\\
\Big(\frac{dR_{S}}{dt}\Big)^{2}&=&1\!-\!\kappa r_{\Sigma}^{2}-F
\label{M2}\\
2\frac{d^{2}R_{S}}{dt^{2}}&=&-\frac{dF}{dR}|_{R_{S}}
\label{M3}
\end{eqnarray}
with $F=1-\frac{2G_k M}{R_S}-\frac{1}{3}\Lambda_k R_S^2$

Eqs. (\ref{M1}, \ref{M2}, \ref{M3}) give the conventional FLRW expansion rate and cosmic acceleration equations for constant $G_{k}$ and $\Lambda_{k}$. For the cse of AS gravity Eqs. (\ref{M1}, \ref{M2}, \ref{M3}) provide modified equations of motion that generate the recent passage from a decceleration to an acceleration era.

AS theory at the present incomplete stage cannot provide the exact dependence of the cutoff $k$ on physical scales like lenght or energy. Thus, for phenomenological purpose various simple scalings have been proposed 
\cite{Bonanno:2000ep}, \cite{Bonanno:2019ilz},
\cite{Bonanno:2007wg}, \cite{2016PhRvD..94j3514K}, \cite{Bonanno:2017pkg}. In the AS swiss cheese $k$ is associated with a characteristic astrophysical length scale $L$. So, the ansatz for $k$ is $k=\xi/L$, with $\xi$ is a dimensionless order-one number. A natural choice which generates the desired phenomenology, is to set as $L$ equal to the proper distance $D>0$.  The use of proper distances proved also to be a sucesful choice also for excibiting a singularity avoidance/smoothing in spherical solutions of AS gravity
\cite{Kofinas:2015sna}.

The proper distance along a radial path, $dT=d\theta=d\varphi=0$ from $R_0$ till $R$ is given by
\begin{equation}
D(R)=\int_{R_{0}}^{R}\frac{d\mathcal{R}}{\sqrt{F}}\,.
\label{proper}
\end{equation}
The value of $k$ is $k_{S}=\xi/D_{S}$, where $D_{S}(R_{S})$ is the proper distance at the matching surface. 
Assuming a far IR fixed point the dimensionless Newton constant and cosmological constant run according to the trajectory
\begin{equation} 
	g\left( k \right) = g_{*} + h_{1}k^{\theta_{1}}, \quad \quad
	\lambda\left( k \right) = \ \lambda_{*} + h_{2}k^{\theta_{2}}, 
\end{equation}
with $(\theta_{1},\theta_{2} \geq 0)$ two unknown critical exponents.

Finally, solving the ODEs \ref{M2},\ref{M3}, that describe the cosmic evolution and refer to the expansion of the Universe we can determine the time evolution of the dark matter contribution, $\Omega_{DM}$, the dark energy contribution $\Omega_{DE}$, the equation of state parameter $w_{DE}$ and the deceleration parameter $q=-\frac{a\ddot{a}}{\dot{a}^2}$.  All the important quantities as functions of the redshift are give n below. 

The derivative with respect to the redshift of the proper distance is 
\begin{equation}
	\frac{D_p}{dz}=-\frac{r_{\Sigma}}{s\,\,(z+1)^2 }
\end{equation}
with 
\begin{equation}
	s=-\frac{2 M (z+1) \left(\frac{g_{*} D_p(z)^2}{\xi ^2}+h_1 \left(\frac{\xi }{D_p(z)}\right)^{\theta_1-2}\right)}{r_{\Sigma}}-\frac{r_{\Sigma}^2 \left(h_2 \left(\frac{\xi }{D_p(z)}\right)^{\theta_2+2}+\frac{\lambda_{*} \xi ^2}{D_p(z)^2}\right)}{3 (z+1)^2}+1
\end{equation}

This is one of the differential equations we have to solve together with the expansion rate ODE for the scale factor.

\begin{equation}
	\begin{split}
		q(z)=&\biggl[ 6 J M D_p(z)^5 \left(h_1 \left(\frac{\xi}{D_p(z)}\right)^{\theta_1}+g_*\right)+\frac{6 \sqrt{3} M r_{\Sigma} D_p(z)^4 \left(h_1 (\theta_1-2) \left(\frac{\xi}{D_p(z)}\right)^{\theta_1}-2 g_*\right)}{z+1} \nonumber \\
		&-\frac{2 J \xi ^4 r_{\Sigma}^3 D_p (z) \left(h_2\left(\frac{\xi}{D_p(z)}\right)^{\theta_2}+\lambda_*\right)}{(z+1)^3}+\frac{\sqrt{3} \xi ^4 r_{\Sigma}^4 \left(h_2 (\theta_2+2) \left(\frac{\xi}{D_p(z)}\right)^{\theta_2}+2 \lambda_*\right)}{(z+1)^4} \biggr] /c
	\end{split}
\end{equation}
where
\begin{equation}
	c=
	{2 y \left(6 M D_p(z)^5 \left(h_1\left(\frac{\xi}{D_p(z)}\right)^{\theta_1}+g_*\right)+\frac{\xi ^4 r_{\Sigma}^3 D_p (z) \left(h_2 \left(\frac{\xi }{D_p(z)}\right)^{\theta_2}+\lambda_*\right)}{(z+1)^3}\right)}
\end{equation}
and
\begin{equation}
	y=\big(-\frac{6 M (z+1) D_p(z)^2 \left(h_1 \left(\frac{\xi }{D_p(z)}\right)^{\theta_1}+g_*\right)}{\xi ^2 r_{\Sigma}}-\frac{\xi ^2 r_{\Sigma}^2 \left(h_2 \left(\frac{\xi }{D_p(z)}\right)^{\theta_2}+\lambda_*\right)}{(z+1)^2 D_p(z)^2}+3\big)^{1/2}
\end{equation}
where the today Schücking radius for a cluster of galaxies with mass $M$ is

\begin{equation}
	r_{\Sigma}=\sqrt[3]{\frac{2 G_N M}{H_0^2 \Omega_{DM\,0}}}
\end{equation}
the dark energy and dark matter part evolution is :

\begin{equation}
	\Omega_{DE}(z) = \frac{ \Omega_{m0} (z+1)^3 \left(\frac{g_* {Dp} (z)^2}{\xi ^2}+h1 \left(\frac{\xi }{{D_p}(z)}\right)^{\theta_ 1-2}\right)} {G_N} +\frac{{h2} \left(\frac{\xi }{Dp(z)}\right)^{\theta_2+2}+\frac{\lambda_* \xi ^2}{Dp(z)^2}}{3\,H_0^2} 
	-\Omega_{m0} (z+1)^3
\end{equation}

\begin{equation}
	\Omega_{DM}(z) = \Omega_{m0} (z+1)^3
\end{equation}

The evolution of the dark energy coefficient is 

\begin{equation}
	\begin{split}
		w_{DE}(z) = & \frac{r_{\Sigma}}{v} \biggl[ \sqrt{3}\, r_{\Sigma}^3\, \xi^4 
		\big( 2 \lambda_*+h_2 (2+\theta_2) (\frac{\xi}{D_p(z)})^{\theta_2} \big) 
		-3 \tilde{J} r_{\Sigma}^2 (1+z) \xi^4 \big[\lambda_*+h_2 (\frac{\xi}{D_p (z)})^{\theta_2}\big] D_p (z) \nonumber \\
		&-6 \sqrt{3} M (1+z)^3 \big(2 g_{*} -h_1 (-2+\theta_1)\left(\frac{\xi}{D_p(z)}\big)^{\theta_1}\right)D_p (z)^4 \biggr]
	\end{split}
\end{equation}
where

\begin{equation}
	v =	3\, j \,(1+z)^4 \biggr[  r_{\Sigma}^3 \xi^4 \big(\lambda_{*} + h_2 \big( \frac{\xi}{D_p (z)} \big) ^{\theta_2} \big) D_p(z)/(1+z)^3 - 6 G_N M \xi^2 D_p (z)^3 + 6 M \big( g_{*} +h_1 \big( \frac{\xi}{D_p(z)}\big)^{\theta_1}\big) D_p(z)^5 \biggl]
\end{equation}

and
\begin{equation}
	j= \sqrt{\left\lbrace 3-\frac{ r_{\Sigma}^2 \xi^2 [\lambda_*+h_2 (\frac{\xi}{D_p(z)})^{\theta_2}]}{(1+z)^2 D_p (z)^2}-\frac{6 M (1+z) [g_*+h_1 (\frac{\xi}{D_p(z)})^{\theta_1}] D_p (z)^2}{r_{\Sigma} \xi^2}\right\rbrace }
\end{equation}
 It has be proven in\cite{Anagnostopoulos:2022pxa} that solving numerically the ODEs \ref{M2},\ref{M3}, the desired behaviour for the dark energy contribution $\Omega_{DE}$ and the equation of state parameter $w_{DE}$ emerges. Most importantly the recent passage from a deceleration to acceleration that happens recently appears.

\section{Modified Einstein equations}

In \cite{Bonanno:2020qfu} authors derived new modified Eistein equations for a gravity with varying $\Lambda$ and $G$ and including up to second order covariant derivatives for the metric the $G$ and $\Lambda$. These extra kinetic terms are absolutely necessary in order to have meaningful energy conservation though Bianchi identities. The derived equations are appropriate to study AS gravity properties.

The gravitational equations demanding the previously mentioned features have been proved in \cite{Bonanno:2020qfu} are:
\begin{equation}
G_{\mu\nu}=-\bar{\Lambda}\,e^{\psi} g_{\mu\nu}
-\frac{1}{2}\psi_{;\mu}\psi_{;\nu}-\frac{1}{4}g_{\mu\nu}\psi^{;\rho}
\psi_{;\rho}+\psi_{;\mu;\nu}-g_{\mu\nu}\Box\psi+8\pi G T_{\mu\nu}
\label{skrvgd}
\end{equation}
and Bianchi identities generate the following  conservation equation
\begin{equation}
\big(GT_{\mu\nu}\big)^{;\mu}\!+\!G\Big(T_{\mu\nu}\!-\!\frac{1}{2}Tg_{\mu\nu}\Big)
\psi^{;\mu}=0\,.
\label{skbmn}
\end{equation}
where $\Lambda=\bar{\Lambda}\,e^{\psi}$ and 
$G=\bar{G}\,e^{\chi}$.

The system of equations (\ref{skrvgd}), (\ref{skbmn}) is unique, for Einstein gravity with varying $\Lambda$ and $G$ allowing up to
second derivatives in $g_{\mu\nu},\Lambda,G$. Note, also that $\Lambda$ and $G$ can be any function of spacetime coordinates. We immediately observe that there are no covariant derivatives for $G$ in equation (\ref{skrvgd}); they cancel out. 
Furthermore, both equations (\ref{skrvgd}), (\ref{skbmn}) reduce to the conventional ones of General Relativity for constant values of $\Lambda$ and $G$.

We re interested to explore the cosmology of these Einstein modified equations. Thus, we work with a spatially homogeneous and isotropic cosmological metric of the form of equation(\ref{eq:FRW}).
To proceed futher we choose a diagonal
energy-momentum tensor $T^{\mu}_{\nu}$, that describes a fluid with energy density
$\rho$ and pressure $P$. It is given by
\begin{equation}
T^{\mu\nu}=(\rho+P) u^{\mu}u^{\nu}+P g^{\mu\nu}\,,
\label{tmn}
\end{equation}
with $u^{\mu}$ the fluid 4-velocity.

Then the conservation equation (\ref{skbmn}) is
\begin{equation}
\dot{\rho}+3nH(\rho+P)+\rho\dot{\chi}+\frac{\rho+3P}{2}\dot{\psi}=0\,.
\label{hyeh}
\end{equation}
We observe that an energy exchange between the energy density $\rho$ and $G,\Lambda$ is allowed.
Since we are interested to describe the late cosmology without any exotic fields or fluids (a "baroque" way to generate dark energy) it suffices to consider a perfect fluid with a constant equation of state parameter
$w=\frac{P}{\rho}$. Then, the conservation equation (\ref{hyeh}) can be written as
\begin{equation}
\Lambda^{\frac{1+3w}{2}}G\rho=\frac{c\,\bar{\Lambda}^{\frac{1+3w}{2}}}{a^{3(1+w)}}\,,
\label{kiyet}
\end{equation}
with $c>0$ an integration constant.

Finally the cosmological equations of motions are
\begin{eqnarray}
&&H^{2}+\frac{\kappa}{a^{2}}=\frac{\bar{\Lambda}}{3}e^{\psi}-H\,\dot{\psi}
-\frac{\dot{\psi}^{2}}{4}+\frac{8\pi c}{3a^{3(1+w)}}\,e^{-\frac{1+3w}{2}\psi}\,,
\label{hywte}\\
&&\dot{H}=\frac{\kappa}{a^{2}}+H\frac{\dot{\psi}}{2}+\frac{\dot{\psi}^{2}}{4}
-\frac{1}{2}\Big(\dot{\psi}\Big)^{\cdot} -\frac{4\pi c (1\!+\!w)}{a^{3(1+w)}}\,e^{-\frac{1+3w}{2}\psi}\,.
\label{hwynqo}
\end{eqnarray}
Note that there is a redundancy. Two out of the three equations (\ref{kiyet}), (\ref{hywte}), (\ref{hwynqo}) are independent. One equation can be derived from the other two, which is a consistent thing to expect. For example equation (\ref{hwynqo}) can be derived from the other two. However, equation (\ref{hwynqo}) is very useful to keep since it provides the acceleration and thus we can check if there is a passage from deceleration to acceleration.

Late cosmic era does not refer to the far infrared regime but in the intermediate IR scale. This can me modeled using the dimensionful quantities, as
\begin{equation}\label{ir1}
{G(k)}_{{IR}} = \frac{g_{*}}{k^{2}} + h_{1}k^{\theta_{1} - 2} \quad\quad
\Lambda\left( k \right)_{{IR}} = \ \lambda_{*}k^{2} + h_{2}k^{\theta_{2}+2} 
\end{equation}
Since we are interested to explore in late cosmology the possibility of solving the dark energy problem without dark energy coming from exotic ad hoc fields and without fine tuning, we can now set $w=0$ and for simplicity $\kappa=0$. Furthermore, to be consistent with solar/galactic dynamics $G$ should be almost constant. Thus, a reasonable choice to proceed is to assume that the second term in the first expression of Eq.(\ref{ir1}) to be dominant to the first. So $G(k)$ behaves as almost constant having a mild dependence of $k$ for appropriate values of $\theta_1$. Thus, $h_1$ is essentially close to $G_N$ and $\theta_1$ must be close to $2$.
Furthermore, for reasonable phenomenology and for reasons arising from AS theory, $\theta_2$ should be close to $0$. Then, both terms in the second expression of Eq.(\ref{ir1}) are almost proportional to $k^2$, we choose for simplicity  $\lambda_{*} \ll 1$.
Therefore, in summary, we assume
\begin{equation}
g_{*} \ll h_{1}k_{IR}^{\theta_{1}} , \quad
\lambda_{*} \ll h_{2}k_{IR}^{\theta_{2}}
\end{equation}
where $k_{IR}$ is the astrophysical scale.

At this stage a certain scale should be selected for $k$. For cosmology it is common to use the following expression \cite{Bonanno:2001xi}
\begin{equation}
k=\frac{\xi}{t}\,,
\label{scale1a}
\end{equation}
where $\xi>0$ is a dimensionless parameter that should be of order of unity in order to avoid a fine tuning problem.

Another popular cutoff identification used in the Asymptotic Safe Gravity phenomenology is a connection with the Hubble scale, \cite{Reuter:2005kb}, 
\begin{equation}
k=\xi H(t)\,,
\label{scale1}
\end{equation}
with $\xi$ a positive constant for $H>0$ which is the case in our investigation. 

For both scaling choices, numerical solutions of equations (\ref{kiyet}), (\ref{hywte}), (\ref{hwynqo}), exhibit no passage from a deceleration to an acceleration era today. This is an expected  and correct result since we have chosen to work with no fine tuning values for the unknown parameters and also we did not include any exotic field or fluid but just a dust content.
The modified Einstein equations presented in \cite{Bonanno:2020qfu} is most probably a fair description of the early Universe assuming that Asymptotic Safe Gravity is a valid model. Furthermore, to explain the passage from deceleration to acceleration, the large scale structure should be taken into account. This could be possible developing a new swiss cheese type model or a new inhomogenous model like Szekeres model for the novel modified Einstein equation (\ref{skrvgd}) respecting equation (\ref{skbmn}).

\section{Conclusions}
\label{III}

In this study we present a brief description of AS gravity and some of its phenomenological consequences regarding Cosmological ans astrophysical issues, with focus on the dark energy problem. Then we reviewed briefly the outcome of a series of publications that propose a natural explanation of recent cosmic acceleration and its coincidence problem with the help of a swiss cheese model. This AS swiss cheese model is consistent with current observations and is well suited to explain dark energy problem in a minimal way without exotic field and without fine tuning or extra scales.

Finally, we have described the late cosmology era of new modified Einstein equations inspired from AS gravity. The analysis of the solutions make apparent that the large scale structure is needed to explain the recent passage from deceleration to acceleration in a natural way.
Thus, as a future work it would be interesting to develop a new swiss cheese model or inhomogeneous models for the new modified Einstein equations.


\begin{thebibliography}{99}

	
	\bibitem{ASreviews}
	M.~Niedermaier and M.~Reuter,
	Living Rev.\ Rel.\  {\bf 9}, 5 (2006);
	R.~Percacci,
	In {\textit{Oriti, D. (ed.): Approaches to quantum gravity}} 111-128
	[arXiv:0709.3851 [hep-th]];
	O.~Lauscher and M.~Reuter,
	In {\textit{Fauser, B. (ed.) et al.: Quantum gravity}} 293-313 [hep-th/0511260];
	M.~Reuter and F.~Saueressig,
	New J.\ Phys.\  {\bf 14}, 055022 (2012)
	[arXiv:1202.2274 [hep-th]];
	A.~Bonanno,
	PoS CLAQG {\bf 08}, 008 (2011) [arXiv:0911.2727 [hep-th]];
	M.~Niedermaier,
	Class.\ Quant.\ Grav.\  {\bf 24}, R171 (2007) [gr-qc/0610018];
	R. Percacci, {\textit{``An Introduction to Covariant Quantum Gravity and Asymptotic Safety''}},
	Word Scientific, ISBN: 978-981-3207-17-2;
	R. Percacci and D. Perini, Phys. Rev. D68, 044018 (2003).
	
	
	\bibitem{Falls:2014tra}
	K.~Falls, D.~F.~Litim, K.~Nikolakopoulos and C.~Rahmede,
	Phys.\ Rev.\ D {\bf 93}, no. 10, 104022 (2016)
	doi:10.1103/PhysRevD.93.104022
	[arXiv:1410.4815 [hep-th]];
	
	
		\bibitem{c11}
	Weinberg, S. Ultraviolet divergences in quantum theories of gravitation. General Relativity: An Einstein centenary
	survey, Eds. Hawking, S.W., Israel, W; Cambridge University Press 1979, pp. 790–831.
	
	\bibitem{c12}
	Reuter, M. Nonperturbative evolution equation for quantum gravity. Phys.Rev. 1998, D57, 971–985,
	[arXiv:hep-th/hep-th/9605030]. doi:10.1103/PhysRevD.57.971.
	
	\bibitem{FRG}
	C. Wetterich, Exact evolution equation for the effective potential, Physics Letters B 301 (1993) 90–94.
	doi:10.1016/0370-2693(93)90726-X.
	; T. R. Morris, The Exact Renormalization Group and Approximate Solutions, International Journal of
	Modern Physics A 9 (1994) 2411–2449. arXiv:hep-ph/9308265, doi:10.1142/S0217751X94000972.
	; M. Reuter, C. Wetterich, Effective average action for gauge theories and exact evolution equations,
	Nuclear Physics B 417 (1994) 181–214. doi:10.1016/0550-3213(94)90543-6.
	
	
	
	
	
\bibitem{Zarikas:2017gfv}
G.~Kofinas and V.~Zarikas,
Phys. Rev. D \textbf{97} (2018) no.12, 123542
doi:10.1103/PhysRevD.97.123542
[arXiv:1706.08779 [gr-qc]].

\bibitem{Anagnostopoulos:2018jdq}
F.~K.~Anagnostopoulos, S.~Basilakos, G.~Kofinas and V.~Zarikas,
``Constraining the Asymptotically Safe Cosmology: cosmic acceleration without dark energy'',
JCAP {\bf 1902}, 053 (2019),
doi:10.1088/1475-7516/2019/02/053
[arXiv:1806.10580 [astro-ph.CO]].

\bibitem{Anagnostopoulos:2019mrc}
F.~K.~Anagnostopoulos, G.~Kofinas and V.~Zarikas,
Int. J. Mod. Phys. D \textbf{28} (2019) no.14, 14
doi:10.1142/S0218271819440139
[arXiv:2102.07578 [gr-qc]].
	
		\bibitem{Anagnostopoulos:2022pxa}
		F.~K.~Anagnostopoulos, A.~Bonanno, A.~Mitra and V.~Zarikas,
		Phys. Rev. D \textbf{105} (2022) no.8, 083532
		doi:10.1103/PhysRevD.105.083532
		[arXiv:2201.02251 [gr-qc]].
		
		\bibitem{Mitra:2021ahd}
		A.~Mitra, V.~Zarikas, A.~Bonanno, M.~Good and E.~G\"udekli,
		Universe \textbf{7} (2021) no.8, 263
		doi:10.3390/universe7080263
		[arXiv:2107.08519 [gr-qc]].


\bibitem{Bonanno:2020qfu}
A.~Bonanno, G.~Kofinas and V.~Zarikas,
Phys. Rev. D \textbf{103} (2021) no.10, 104025
doi:10.1103/PhysRevD.103.104025
[arXiv:2012.05338 [gr-qc]].


	\bibitem{cosmoreview}	
		A. Bonanno, F. Saueressig, Asymptotically safe cosmology - A status report, Comptes Rendus
	Physique 18 (2017) 254–264. arXiv:1702.04137, doi:10.1016/j.crhy.2017.02.002.
	; A.~Platania,
	Front. in Phys. \textbf{8} (2020), 188
	doi:10.3389/fphy.2020.00188
	[arXiv:2003.13656 [gr-qc]].
		
\bibitem{Kofinas:2015sna}
G.~Kofinas and V.~Zarikas,
JCAP \textbf{10} (2015), 069
doi:10.1088/1475-7516/2015/10/069
[arXiv:1506.02965 [hep-th]].

\bibitem{Kofinas:2016lcz}
G.~Kofinas and V.~Zarikas,
Phys. Rev. D \textbf{94} (2016) no.10, 103514
doi:10.1103/PhysRevD.94.103514
[arXiv:1605.02241 [gr-qc]].




\bibitem{Zarikas:2018wfv}
V.~Zarikas and G.~Kofinas,
J. Phys. Conf. Ser. \textbf{1051} (2018) no.1, 012028
doi:10.1088/1742-6596/1051/1/012028
[arXiv:2006.08674 [gr-qc]].
  
	\bibitem{Reuter:2012xf}
	M.~Reuter, F.~Saueressig,
	Lect.\ 	Notes Phys.\ 863 (2013) 185,
	\newblock \href {http://arxiv.org/abs/1205.5431} {\path{arXiv:1205.5431}}.
	
	\bibitem{Bonanno:2001xi}
	A.~Bonanno, M.~Reuter,
	Phys.\ Rev.\ D65 (2002) 043508,
	\newblock \href {http://arxiv.org/abs/hep-th/0106133}
	{\path{arXiv:hep-th/0106133}}.
	
	\bibitem{Bonanno:2001hi}
	A.~Bonanno, M.~Reuter, 
	 Phys.\ Lett.\ B527 (2002) 9,
	\newblock \href {http://arxiv.org/abs/astro-ph/0106468}
	{\path{arXiv:astro-ph/0106468}}.
	
	\bibitem{Bentivegna:2003rr}
	E.~Bentivegna, A.~Bonanno, M.~Reuter,
	 JCAP 01 (2004) 001,
	\newblock \href {http://arxiv.org/abs/astro-ph/0303150}
	{\path{arXiv:astro-ph/0303150}}. 
	
	\bibitem{Reuter:2005kb}
	M.~Reuter, F.~Saueressig, 
	 JCAP 09 (2005) 012,
	\newblock \href {http://arxiv.org/abs/hep-th/0507167}
	{\path{arXiv:hep-th/0507167}}. 
	
	\bibitem{Bonanno:2007wg}
	A.~Bonanno, M.~Reuter, 
	JCAP 08 (2007) 024,
	\newblock \href {http://arxiv.org/abs/0706.0174} {\path{arXiv:0706.0174}}.
	
	\bibitem{Weinberg:2009wa}
	S.~Weinberg, 
	 Phys.\ Rev.\ D81 (2010) 083535,
	\newblock \href {http://arxiv.org/abs/0911.3165} {\path{arXiv:0911.3165}}.
	
	\bibitem{Bonanno:2009nj}
	A.~Bonanno, 
	PoS CLAQG08 (2011) 008,
	\newblock \href {http://arxiv.org/abs/0911.2727} {\path{arXiv:0911.2727}}.
	
	\bibitem{Bonanno:2010mk}
	A.~Bonanno, M.~Reuter, 
	Entropy 13 (2011) 274,
	\newblock \href {http://arxiv.org/abs/1011.2794} {\path{arXiv:1011.2794}}.
	
	\bibitem{Koch:2010nn}
	B.~Koch, I.~Ramirez,
	 Class.\ Quant.\ Grav.\ 28 (2011) 055008,
	\newblock \href {http://arxiv.org/abs/1010.2799} {\path{arXiv:1010.2799}}.
	
	\bibitem{Casadio:2010fw}
	R.~Casadio, S.~D.~H. Hsu, B.~Mirza, 
	Phys.\ Lett.\ B695 (2011) 317, 
	\newblock \href {http://arxiv.org/abs/1008.2768} {\path{arXiv:1008.2768}}.
	
	\bibitem{Contillo:2010ju}
	A.~Contillo, 
	 Phys.\ Rev.\ D83 (2011) 085016,
	\newblock \href {http://arxiv.org/abs/1011.4618} {\path{arXiv:1011.4618}}.
	
	\bibitem{Bonanno:2010bt}
	A.~Bonanno, A.~Contillo, R.~Percacci,
	Class.\ Quant.\ Grav.\ 28 (2011) 145026,
	\newblock \href {http://arxiv.org/abs/1006.0192} {\path{arXiv:1006.0192}}.
	
	\bibitem{Frolov:2011ys}
	A.~V.~Frolov, J.-Q.~Guo, 
	\href {http://arxiv.org/abs/1101.4995}
	{\path{arXiv:1101.4995}}.
	
	\bibitem{Hindmarsh:2011hx}
	M.~Hindmarsh, D.~Litim, C.~Rahmede,
	JCAP 07 (2011) 019,
	\newblock \href {http://arxiv.org/abs/1101.5401} {\path{arXiv:1101.5401}}.
	
	\bibitem{Bonanno:2011yx}
	A.~Bonanno, S.~Carloni, 
	New J.\ Phys.\ 14 (2012) 025008,
	\newblock \href {http://arxiv.org/abs/1112.4613} {\path{arXiv:1112.4613}}.
	
	\bibitem{Ahn:2011qt}
	C.~Ahn, C.~Kim, E.~V. Linder, 
	Phys.\ Lett.\ B704 (2011) 10,
	\newblock \href {http://arxiv.org/abs/1106.1435} {\path{arXiv:1106.1435}}.
	
	\bibitem{Cai:2011kd}
	Y.-F.~Cai, D.~A.~Easson,
	 Phys.\ Rev.\ D84 (2011) 103502,
	\newblock \href {http://arxiv.org/abs/1107.5815} {\path{arXiv:1107.5815}}.
	
	\bibitem{Contillo:2011ag}
	A.~Contillo, M.~Hindmarsh, C.~Rahmede, 
	 Phys. Rev. D85 (2012) 043501,
	\newblock \href {http://arxiv.org/abs/1108.0422} {\path{arXiv:1108.0422}}.
	
	\bibitem{Cai:2012qi}
	Y.-F.~Cai, D.~A.~Easson,
	Int.\ J.\ Mod.\ Phys.\ D21 (2013) 1250094,
	\newblock \href {http://arxiv.org/abs/1202.1285} {\path{arXiv:1202.1285}}.
	
	\bibitem{Bonanno:2012jy}
	A.~Bonanno, 
	Phys.\ Rev.\ D85 (2012) 081503,
	\newblock \href {http://arxiv.org/abs/1203.1962} {\path{arXiv:1203.1962}}.
	
	\bibitem{Hindmarsh:2012rc}
	M.~Hindmarsh, I.~D. Saltas, 
	Phys.\ Rev.\ D86 (2012) 064029,
	\newblock \href {http://arxiv.org/abs/1203.3957} {\path{arXiv:1203.3957}}.
	
	\bibitem{Fang:2012ca}
	C.~Fang, Q.-G. Huang, 
	Eur.\ 	Phys.\ J.\ C73 (2013) 2401,
	\newblock \href {http://arxiv.org/abs/1210.7596} {\path{arXiv:1210.7596}}.
	
	\bibitem{Bonanno:2013dja}
	A.~Bonanno, M.~Reuter, 
	Phys.\ Rev.\ D87 (2013) 084019,
	\newblock \href {http://arxiv.org/abs/1302.2928} {\path{arXiv:1302.2928}}.
	
	\bibitem{Copeland:2013vva}
	E.~J. Copeland, C.~Rahmede, I.~D. Saltas,
	Phys.\ Rev.\ D91 (2015) 103530,
	\newblock \href {http://arxiv.org/abs/1311.0881} {\path{arXiv:1311.0881}}.
	
	\bibitem{Kaya:2013bga}
	A.~Kaya, 
	Phys.\ Rev.\ D87 (2013) 123501,
	\newblock \href {http://arxiv.org/abs/1303.5459} {\path{arXiv:1303.5459}}.
	
	\bibitem{Becker:2014jua}
	D.~Becker, M.~Reuter, 
	JHEP 12 (2014) 025,
	\newblock \href {http://arxiv.org/abs/1407.5848} {\path{arXiv:1407.5848}}.
	
	\bibitem{Xianyu:2014eba}
	Z.-Z. Xianyu, H.-J. He,
	JCAP 10 (2014) 083,
	\newblock \href {http://arxiv.org/abs/1407.6993} {\path{arXiv:1407.6993}}.
	
	\bibitem{Saltas:2015vsc}
	I.~D. Saltas, 
	JCAP 02 (2016) 048,
	\newblock \href {http://arxiv.org/abs/1512.06134} {\path{arXiv:1512.06134}}.
	
	\bibitem{Nielsen:2015una}
	N.~G. Nielsen, F.~Sannino, O.~Svendsen, 
	Phys.\ Rev.\ D91 (2015) 103521,
	\newblock \href {http://arxiv.org/abs/1503.00702} {\path{arXiv:1503.00702}}.
	
	\bibitem{Bonanno:2015fga}
	A.~Bonanno, A.~Platania, 
	Phys.\ Lett.\ B750 (2015) 638, 
	\newblock \href {http://arxiv.org/abs/1507.03375} {\path{arXiv:1507.03375}}.
	

\bibitem{Eichhorn:2022bgu}
A.~Eichhorn and A.~Held,
[arXiv:2212.09495 [gr-qc]].


\bibitem{Bonanno:2017zen}
A.~Bonanno, B.~Koch and A.~Platania,
Found. Phys. \textbf{48} (2018) no.10, 1393-1406
doi:10.1007/s10701-018-0195-7
[arXiv:1710.10845 [gr-qc]].


\bibitem{Saueressig:2015xua}
F.~Saueressig, N.~Alkofer, G.~D'Odorico and F.~Vidotto,
PoS \textbf{FFP14} (2016), 174
doi:10.22323/1.224.0174
[arXiv:1503.06472 [hep-th]].


\bibitem{Cai:2010zh}
Y.~F.~Cai and D.~A.~Easson,
JCAP \textbf{09} (2010), 002
doi:10.1088/1475-7516/2010/09/002
[arXiv:1007.1317 [hep-th]].

\bibitem{Falls:2010he}
K.~Falls, D.~F.~Litim and A.~Raghuraman,
Int. J. Mod. Phys. A \textbf{27} (2012), 1250019
doi:10.1142/S0217751X12500194
[arXiv:1002.0260 [hep-th]].




\bibitem{Gubitosi:2018gsl}
G.~Gubitosi, R.~Ooijer, C.~Ripken and F.~Saueressig,
JCAP \textbf{12} (2018), 004
doi:10.1088/1475-7516/2018/12/004
[arXiv:1806.10147 [hep-th]].


\bibitem{Lehners:2019ibe}
J.~L.~Lehners and K.~S.~Stelle,
Phys. Rev. D \textbf{100} (2019) no.8, 083540
doi:10.1103/PhysRevD.100.083540
[arXiv:1909.01169 [hep-th]].


  
	\bibitem{Zarikas:2011pq} 
	G.~Kofinas and V.~Zarikas,
	Eur.\ Phys.\ J.\ C {\bf 73}, no. 4, 2379 (2013)
	doi:10.1140/epjc/s10052-013-2379-9
	[arXiv:1107.2602 [hep-th]].
	
	

	
\bibitem{Li:2011sd}
M.~Li, X.~D.~Li, S.~Wang and Y.~Wang,
Commun. Theor. Phys. \textbf{56} (2011), 525-604
doi:10.1088/0253-6102/56/3/24
[arXiv:1103.5870 [astro-ph.CO]].
	
	\bibitem{IR}
	A.~Bonanno and M.~Reuter,
	Phys. Lett. B \textbf{527} (2002), 9-17
	doi:10.1016/S0370-2693(01)01522-2
	[arXiv:astro-ph/0106468 [astro-ph]].
	M.~Reuter and H.~Weyer,
	JCAP \textbf{12} (2004), 001
	doi:10.1088/1475-7516/2004/12/001
	[arXiv:hep-th/0410119 [hep-th]].
	G.~Esposito, C.~Rubano and P.~Scudellaro,
	Class. Quant. Grav. \textbf{24} (2007), 6255-6266
	doi:10.1088/0264-9381/24/24/008
	[arXiv:0709.1403 [gr-qc]].
	
	
	\bibitem{Koch:2014cqa}
	B.~Koch and F.~Saueressig,
	Int. J. Mod. Phys. A \textbf{29} (2014) no.8, 1430011
	doi:10.1142/S0217751X14300117
	[arXiv:1401.4452 [hep-th]].
	
	
	
		\bibitem{Einstein:1946ev}
	A.~Einstein and E.~G.~Strauss,
	Annals Math.\  {\bf 47}, 731 (1946)
	doi:10.2307/1969231.
	
	
	\bibitem{Israel:1966rt}
	W.~Israel,
	Nuovo Cim.\ B {\bf 44S10}, 1 (1966)
	[Nuovo Cim.\ B {\bf 44}, 1 (1966)]
	Erratum: [Nuovo Cim.\ B {\bf 48}, 463 (1967)]
	doi:10.1007/BF02710419, 10.1007/BF02712210;
	G. Darmois, {\em M\'{e}morial des Sciences Math\'{e}matiques\/},
	Fascicule XXV (Gauthier-Villars, Paris, 1927), Chap. V.
	
	\bibitem{matching} 
	G. Darmois, {\em M\'{e}morial des Sciences Math\'{e}matiques\/},
	Fascicule XXV (Gauthier-Villars, Paris, 1927), Chap. V.;
	H. Stephani, {\em General Relativity\/} (Cambridge University
	Press, Cambridge, 1990);
	G.A. Baker, Jr. {\em Bound systems in an expanding universe\/},
	astro-ph/0003152, 10 March 2000;
	L.P. Eisenhart, {\em Riemannian Geometry\/} (Princeton
	University Press, Princeton, 1949).
	
	
	
	
	
	
	
	\bibitem{Bonanno:2000ep}
	A.~Bonanno and M.~Reuter,
	Phys. Rev. D \textbf{62}, 043008 (2000)
	doi:10.1103/PhysRevD.62.043008
	[arXiv:hep-th/0002196 [hep-th]].
	
	\bibitem{Bonanno:2019ilz}
	A.~Bonanno, R.~Casadio and A.~Platania,
	JCAP \textbf{01}, 022 (2020)
	doi:10.1088/1475-7516/2020/01/022
	[arXiv:1910.11393 [gr-qc]].
	
	
	\bibitem{2016PhRvD..94j3514K}
	G.~Kofinas and V.~Zarikas,
	Phys. Rev. D \textbf{94}, no.10, 103514 (2016)
	doi:10.1103/PhysRevD.94.103514
	[arXiv:1605.02241 [gr-qc]].
	
	\bibitem{Bonanno:2017pkg}
	A.~Bonanno and F.~Saueressig,
	Comptes Rendus Physique \textbf{18}, 254-264 (2017)
	doi:10.1016/j.crhy.2017.02.002
	[arXiv:1702.04137 [hep-th]].
	
	
	
	
	
	
	
	\bibitem{inho}
	G.~F.~R.~Ellis and W.~Stoeger,
	Class.\ Quant.\ Grav.\  {\bf 4}, 1697 (1987)
	doi:10.1088/0264-9381/4/6/025; 
	S.~R.~Green and R.~M.~Wald,
	Phys.\ Rev.\ D {\bf 83}, 084020 (2011)
	doi:10.1103/PhysRevD.83.084020
	[arXiv:1011.4920 [gr-qc]]; 
	T.~Buchert, M.~J.~France and F.~Steiner,
	Class.\ Quant.\ Grav.\  {\bf 34}, no. 9, 094002 (2017),
	doi:10.1088/1361-6382/aa5ce2
	[arXiv:1701.03347 [astro-ph.CO]]; 
	T.~Buchert, A.~A.~Coley, H.~Kleinert, B.~F.~Roukema and D.~L.~Wiltshire,
	Int.\ J.\ Mod.\ Phys.\ D {\bf 25}, no. 03, 1630007 (2016),
	doi:10.1142/S021827181630007X, 10.1142/9789813226609-0034
	[arXiv:1512.03313 [astro-ph.CO]].
	
	\bibitem{inhoswiss}
	S.~M.~Koksbang,
	Phys.\ Rev.\ D {\bf 95}, no. 6, 063532 (2017)
	doi:10.1103/PhysRevD.95.063532
	[arXiv:1703.03572 [astro-ph.CO]];
	K.~Bolejko and M.~N.~Celerier,
	Phys.\ Rev.\ D {\bf 82}, 103510 (2010)
	doi:10.1103/PhysRevD.82.103510
	[arXiv:1005.2584 [astro-ph.CO]];
	P.~Mishra, M.~N.~Celerier and T.~P.~Singh,
	Phys.\ Rev.\ D {\bf 86}, 083520 (2012)
	doi:10.1103/PhysRevD.86.083520
	[arXiv:1206.6026 [astro-ph.CO]];
	T.~Biswas and A.~Notari,
	JCAP {\bf 0806}, 021 (2008)
	doi:10.1088/1475-7516/2008/06/021
	[astro-ph/0702555];
	V.~Marra, E.~W.~Kolb and S.~Matarrese,
	Phys.\ Rev.\ D {\bf 77}, 023003 (2008)
	doi:10.1103/PhysRevD.77.023003
	[arXiv:0710.5505 [astro-ph]].
	
	\bibitem{structure}
	S.~Rasanen,
	EAS Publ.\ Ser.\  {\bf 36}, 63 (2009)
	doi:10.1051/eas/0936008
	[arXiv:0811.2364 [astro-ph]];
	S.~Rasanen,
	arXiv:1012.0784 [astro-ph.CO];
	R.~A.~Sussman,
	Class.\ Quant.\ Grav.\  {\bf 28}, 235002 (2011)
	doi:10.1088/0264-9381/28/23/235002
	[arXiv:1102.2663 [gr-qc]];
	M.~Lavinto, S.~Rasanen and S.~J.~Szybka,
	JCAP {\bf 1312}, 051 (2013)
	doi:10.1088/1475-7516/2013/12/051
	[arXiv:1308.6731 [astro-ph.CO]].
	
	
	
	
	
	

	\bibitem{c2}
	A.~Bonanno, A.~Eichhorn, H.~Gies, J.~M.~Pawlowski, R.~Percacci, M.~Reuter, F.~Saueressig and G.~P.~Vacca,
	Front. in Phys. \textbf{8} (2020), 269
	doi:10.3389/fphy.2020.00269
	[arXiv:2004.06810 [gr-qc]].
	
	
	\bibitem{c3}
	A.~M.~Polyakov,
	[arXiv:hep-th/9304146 [hep-th]].
	
	\bibitem{c4}
	A.~Bonanno, G.~Kofinas and V.~Zarikas,
	Phys. Rev. D \textbf{103} (2021) no.10, 104025
	doi:10.1103/PhysRevD.103.104025
	[arXiv:2012.05338 [gr-qc]].
	;
	M.~Reuter and H.~Weyer,
	JCAP {\bf 0412}, 001 (2004)
	doi:10.1088/1475-7516/2004/12/001
	[hep-th/0410119];
	;
	M.~Reuter and H.~Weyer,
	Phys.\ Rev.\ D {\bf 70}, 124028 (2004)
	doi:10.1103/PhysRevD.70.124028
	[hep-th/0410117];
	;
	M.~Reuter and H.~Weyer,
	Phys.\ Rev.\ D {\bf 69}, 104022 (2004)
	doi:10.1103/PhysRevD.69.104022
	[hep-th/0311196].
	







\end{thebibliography}
\end{document}